\documentclass[sigconf]{acmart}

\usepackage[ruled,linesnumbered,noend]{algorithm2e}
\usepackage{algpseudocode}
\usepackage{multicol}
\usepackage{multirow}
\usepackage[normalem]{ulem}
\usepackage{amsmath}
\algtext*{EndWhile}
\newcommand{\method}{\texttt{SAD}}
\usepackage{wrapfig}
\usepackage{ulem}
\usepackage{enumitem}
\usepackage{booktabs}
\usepackage{array}
\usepackage{pifont}
\usepackage{colortbl} 
\AtBeginDocument{%
  }

\setcopyright{acmlicensed}
\copyrightyear{2025}
\acmYear{2025}
\acmDOI{XXXXXXX.XXXXXXX}

\acmConference[EASE 2025]{The 29th International Conference on Evaluation and Assessment in Software Engineering}{17–20 June, 2025}{Istanbul, Türkiye}
\acmISBN{978-1-4503-XXXX-X/18/06}

\begin{document}


\title{Version-level Third-Party Library Detection in Android Applications via Class Structural Similarity}


\newcommand{\inst}[1]{\textsuperscript{#1}} 

\author{Bolin Zhou}
\affiliation{
    \institution{Institute of Software, Chinese Academy of Sciences}
    \city{} \country{}
}
\affiliation{
    \institution{University of Chinese Academy of Sciences}
    \city{} \country{}
}
\email{zhoubolin22@mails.ucas.ac.cn}


\author{Jingzheng Wu}
\affiliation{%
  \institution{Institute of Software, Chinese Academy of Sciences}
  \city{}
  \country{}
  }
\affiliation{%
  \institution{Key Laboratory of System Software (Chinese Academy of Sciences)}
  \city{}
  \country{}
  }
\affiliation{%
  \institution{State Key Laboratory of Computer Science, Institute of Software, Chinese Academy of Sciences}
  \city{}
  \country{}
  }
\email{jingzheng08@iscas.ac.cn}

\author{Xiang Ling}
\authornote{\ding{41} Xiang Ling is the corresponding author.}
\affiliation{
  \institution{Institute of Software, Chinese Academy of Sciences}
  \city{}
  \country{}
  }
\affiliation{%
  \institution{Key Laboratory of System Software (Chinese Academy of Sciences)}
  \city{}
  \country{}
  }
\affiliation{%
  \institution{State Key Laboratory of Computer Science, Institute of Software, Chinese Academy of Sciences}
  \city{}
  \country{}
  }
\email{lingxiang@iscas.ac.cn}

\author{Tianyue Luo}
\affiliation{%
  \institution{Institute of Software, Chinese Academy of Sciences}
  \city{}
  \country{}
  }
\email{tianyue@iscas.ac.cn}

\author{Jingkun Zhang}
\affiliation{
    \institution{Institute of Software, Chinese Academy of Sciences}
    \city{} \country{}
}
\email{zhangjingkun23@mails.ucas.ac.cn}
\begin{abstract}
Android applications (apps) integrate reusable and well-tested third-party libraries (TPLs) to enhance functionality and shorten development cycles. However, recent research reveals that TPLs have become the largest attack surface for Android apps, where the use of insecure TPLs can compromise both developer and user interests. To mitigate such threats, researchers have proposed various tools to detect TPLs used by apps, supporting further security analyses such as vulnerable TPLs identification. 

Although existing tools achieve notable library-level TPL detection performance in the presence of obfuscation, they struggle with version-level TPL detection due to a lack of sensitivity to differences between versions. This limitation results in a high version-level false positive rate, significantly increasing the manual workload for security analysts. To resolve this issue, we propose {\method}, a TPL detection tool with high version-level detection performance. {\method} generates a candidate app class list for each TPL class based on the feature of nodes in class dependency graphs (CDGs). It then identifies the unique corresponding app class for each TPL class by performing class matching based on the similarity of their class summaries. Finally, {\method} identifies TPL versions by evaluating the structural similarity of the sub-graph formed by matched classes within the CDGs of the TPL and the app.
Extensive evaluation on three datasets demonstrates the effectiveness of {\method} and its components. {\method} achieves F1 scores of 97.64\% and 84.82\% for library-level and version-level detection on obfuscated apps, respectively, surpassing existing state-of-the-art tools. The version-level false positives reported by the best tool is 1.61 times that of {\method}.
We further evaluate the degree to which TPLs identified by detection tools correspond to actual TPL classes. Experimental results show that {\method} achieves a class-level F1 score of 94.12\%, 11\% higher than the best tool, demonstrating the reliability of {\method} and better supporting downstream tasks that rely on specific code.


\end{abstract}



\received{31 January 2025}
\received[accepted]{22 March 2025}

\maketitle

\section{Introduction}\label{sec:intro}
Third-party libraries (TPLs) in Android offer a wide range of pre-implemented functionalities, enabling developers to avoid reinventing the wheel~\cite{wukong}. This greatly simplifies the development process for Android apps and significantly shortens the app development and delivery cycle~\cite{Viennot_Garcia_Nieh_2014}. However, research~\cite{180236,meng2016price,tpl_are_one_of_the_most_insecure_parts_of_an_application} indicates that TPLs have become the largest attack surface within apps, and the integration of numerous TPLs introduces various security and compliance issues. Specifically, when TPLs containing vulnerabilities, malicious code, or license conflicts with other TPLs are integrated into apps, they may harm developers' interests and jeopardize user privacy and security.
Many studies have been proposed to evaluate and mitigate these threats. For instance, some researchers propose reducing risks posed by TPLs through permission downgrading~\cite{HAMMAD201983,7930200} or process isolation~\cite{180236,10.1145/2523649.2523652}. Wu et al.~\cite{10.1145/3597503.3639132} employ function summaries to detect flaws in TPLs. Other works focus on extracting TPLs to identify and analyze malicious behaviors~\cite{10.1145/3366423.3380242,10.1145/3308558.3313549,7546512,184453}, analyzing fraudulent activities in ad TPLs~\cite{10.1145/3236024.3236045,10.1145/3177102.3177113}, examining privacy leakage issues associated with TPLs~\cite{meng2016price,8952393}, and identifying and analyzing vulnerable TPLs~\cite{7962351,libscout}.

A \textbf{prerequisite task} for aforementioned works is detecting TPLs in apps. The TPL detection supports downstream reliable and effective security and compliance analysis by identifying the TPLs and their versions used in apps~\cite{libscout,orlis,atvhunter,libscan,libloom,libhunter}. Due to the importance of TPL detection, it has become a core technology in many commercial software composition analysis (SCA) products~\cite{appsweep,scantist,huaweiyunsca}.
Nevertheless, although existing advanced TPL detection tools~\cite{libscan, libloom, libpecker, libhunter} achieve library-level F1 scores exceeding 95\%, their performance in version-level detection, which is crucial for many downstream tasks, remains suboptimal. This can be attributed primarily to the following two reasons.
\textbf{R\#1: Reliance on features susceptible to obfuscation.} Existing tools rely on code features that are vulnerable to obfuscation, making it more challenging to detect TPLs in obfuscated apps. This leads to more false negatives and degrades detection recall at both the library- and version-level. 
\textbf{R\#2: Unable to capture differences across versions.} Existing TPL detection tools independently process each TPL without capturing the differences between different versions of TPLs, which leads to scenarios where, at the version-level, different versions are detected as being within the app, particularly when the differences are minimal, resulting in a significant number of version-level false positives. 
Notably, although existing studies~\cite{libscan, libhunter} have evaluated the version-level detection performance of TPL detection tools, we argue that their precision calculation methods are overly relaxed and fail to effectively reflect a more realistic version-level detection performance.

In order to address the aforementioned issues, we propose {\method}, which utilizes class structural similairty and summaries of class functionality for the TPL detection. 
Specifically, to deal with issue \textbf{R\#1}, {\method} first leverages the similarity of node features, integrating CDG structural information, to construct a candidate app class list for each TPL class. It then generates class functionality summaries based on field operations for class matching.
Regarding issue \textbf{R\#2}, {\method} extracts fine-grained features of different TPL versions, filters erroneous versions based on the number of high-confidence matched classes, and conducts cross-version comparisons to identify subtle differences, ultimately determining the precise version.

Specifically, {\method} first extracts class-level features and inter-class relationships from the input TPL and app to construct CDGs, and then generates feature for each class node through the operation of neighbor feature aggregation. Then, {\method} generates a candidate app class list for every TPL class by calculating the similarity between the feature of nodes in the CDGs of the TPL and app.
Subsequently, {\method} performs reliable class matching based on the candidate lists of TPL classes. {\method} categorizes classes into two types: \textit{stateful classes} and \textit{stateless classes}, referring to classes with and without non-static fields, respectively. Given a pair of classes from the TPL and the app, {\method} requires their types to be consistent, categorizing them into stateful or stateless pairs. For stateful pairs, {\method} establishes correspondences between the methods and fields of TPL and app classes, then summarizes class functionality by translating lengthy method code into field operation representations, leveraging intra-class method invocation sequences. Class matching is then conducted based on the similarity of these summaries. For stateless pairs, {\method} matches methods using field read/write operations and method opcodes, and performs class matching based on opcodes matching ratio. 
Finally, {\method} determines the TPL version based on the structural similarity of the sub-graph formed by the matched nodes within the CDGs associated with the TPL and the app.

We evaluated the effectiveness of {\method} on three datasets composed of 1123 apps, and 562 TPLs. The results show that {\method} outperforms existing tools on obfuscated dataset at both the library- and version-level~\cite{libscan}, achieving F1 scores of 97.64\% and 95.35\%, respectively. Particularly, {\method} scores 84.82\% in version$^{\dagger}$-level that we propose, which is 10\% higher than the best TPL detection tool. 
Overall, our contributions are as follows:

\begin{itemize}[leftmargin=*]
\item We propose a version-level TPL detection tool {\method}, utilizing class structural similarity between TPLs and apps for TPL detection.
\item We propose a stricter calculation method for version-level false positives. This adjustment reduces the overestimation of F1 scores and provides a more realistic measurement of tool effectiveness.
\item We evaluate the TPL detection performance of different tools on three datasets. {\method} achieves 97.64\% accuracy at the librar-level, comparable to state-of-the-art tools. For version-level detection, it achieves the highest F1 score of 84.82\% among all tools.
\end{itemize}
\section{Background and Related Work}

\subsection{Code Obfuscation}
Code obfuscation techniques are widely employed to protect app code from threats such as piracy and reverse engineering. By applying semantically equivalent transformations, these techniques make code more difficult to understand and analyze, thereby achieving the goal of safeguarding apps.
However, the widespread use of obfuscation techniques poses significant challenges to TPL detection task, making external security audits and research more difficult. As a result, researchers have proposed various methods for detecting obfuscation to guide further analysis. IREA~\cite{7345285} is a tool that employs pattern matching and API recognition to statically detect obfuscations, including identifier renaming, reflection, and data encryption. Mirzaei et al.~\cite{MIRZAEI2019240} introduced AndrODet, which defines heuristic rules for statically detecting three types of obfuscation: identifier renaming, string encryption, and control flow obfuscation. Jiang et al.~\cite{CONTI2022103311} extract control flow graphs (CFGs) and leverage a hybrid model combining graph convolutional network and long short-term memory network to detect string encryption, identifier renaming, and control flow obfuscation at the function level.
In addition to obfuscation detection, some studies have focused on directly deobfuscating obfuscated apps. DeGuard~\cite{deguard} is a tool designed to address identifier renaming obfuscation by learning a probabilistic model from thousands of unobfuscated apps and applying the model to restore identifier names in obfuscated apps. Yoo et al.~\cite{yoo2016string} proposed a dynamic code extraction-based approach to retrieve decrypted strings from apps affected by string encryption obfuscation.

Given the limited scope of existing obfuscation detection and deobfuscation methods, which restrict their ability to serve as preprocessing steps for TPL detection, an increasing number of TPL detection tools have been designed to account for the impact of code obfuscation~\cite{libloom,libscan,libhunter,atvhunter}. Notably, as technology advances, novel obfuscation techniques continue to emerge~\cite{obfuscapk}, making it highly challenging to develop a tool that is resilient to all obfuscation techniques. Therefore, we primarily focus on commonly used obfuscation techniques in widely adopted obfuscation tools~\cite{allatori,proguard,dasho}.

\subsection{TPL Detection Tools}

Existing TPL detection tools can be mainly categorized into two types based on their methodologies~\cite{tplreview}: machine learning-based~\cite{libradar,libD}, and similarity comparison-based~\cite{libscan,atvhunter,libhunter,libloom,libpecker}. 
Tools based on machine learning can be further categorized into classification-based and clustering-based tools. Classification-based tools are primarily designed to distinguish between ad TPLs and non-ad TPLs, but their applicability is limited.
Clustering-based tools utilize a large number of apps as input to extract code features for clustering purposes, grouping similar TPLs together. The underlying principle is that popular TPLs are utilized by numerous apps. However, the limitation of such methods lies in the necessity of acquiring a substantial number of apps, which consequently restricts detection to widely used TPLs, thereby neglecting newer or less common ones. 
The majority of tools are feature similarity comparison-based, which do not require a large number of apps. Instead, they necessitate the construction of a feature database for TPLs, extracting features from the input apps and comparing them with those in the database to identify TPLs within the apps. This method requires pairwise comparisons, which is often time-consuming, yet it offers superior detection performance.

LibScan~\cite{libscan} builds potential class correspondences through fingerprint code features and then detects in-app TPLs using two fine-grained stages of method opcodes similarity and call chain opcode similarity, achieving excellent efficiency and effectiveness. 
LibHunter~\cite{libhunter}, building upon the foundation of LibScan, takes into account the effects of optimizations.
It utilizes an enhanced class signature matching approach to address call site optimizations for constructing class correspondence relationships. Subsequently, method matching is performed using the opcodes and strings of the methods, and method inlining optimizations are processed through simulating method inlining strategies. It has demonstrated outstanding performance on optimized apps.
LIBLOOM~\cite{libloom} converts TPL detection into a set inclusion problem, using two-stage bloom filter to extract candidate TPLs and compute similarity scores for detection, and employs a novel entropy-based metric to specifically handle apps obfuscated by repackaging and package flattening, significantly improving scalability while ensuring effectiveness.
LibPecker~\cite{libpecker} matches TPLs by generating signatures based on class dependencies for both TPL classes and app classes and introduces adaptive class similarity thresholds and weighted class similarity scores when computing TPL similarities, which makes it more resilient to obfuscations. 

Although these tools have been proven to be resilient to obfuscations, the evaluation metrics at the version-level are overly relaxed, leading to overestimated performance of TPL detection tools. As a result, their performance in downstream tasks reliant on specific TPL versions is suboptimal. This motivates the development of a TPL detection tool with superior performance under more accurate and stricter version-level detection metric, thereby supporting a broader range of downstream tasks.

\section{Methodology}
\begin{figure*}[htbp]
    \centering
    \includegraphics[width=\textwidth]{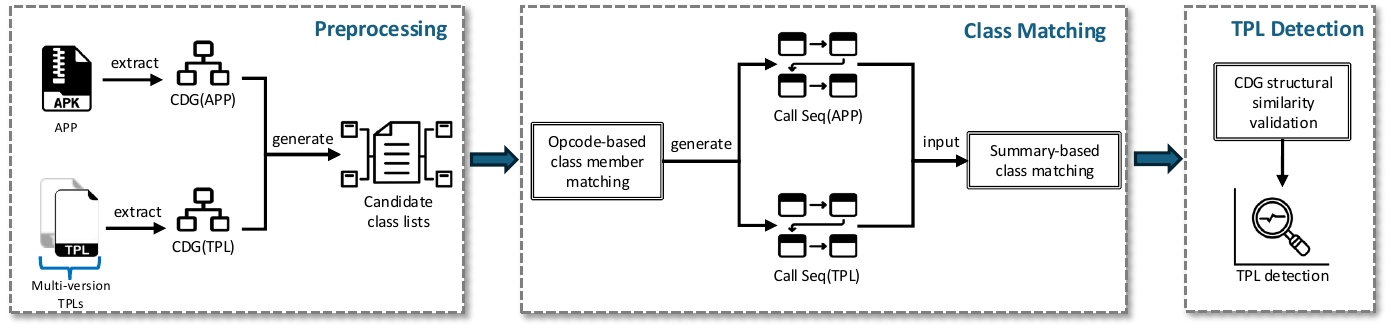}
    \caption{The overview of {\method}, which includes three main modules, preprocessing, class matching and TPL detection.}
    \label{fig:overview}
\end{figure*}

In this section, we describe the functionality of the various modules of {\method}. As shown in Figure~\ref{fig:overview}, the overall framework of {\method} consists of three modules. In the preprocessing module, {\method} first parses the app and the TPL to extract their CDGs. It then constructs a list of candidate app classes for each TPL class based on the feature of nodes in the CDGs associated with the TPL and the app. Subsequently, {\method} performs class matching by iterating over the candidate list for each TPL class using the similarity of class functionality summary. Finally, {\method} validates the structural similairty of the sub-graph formed by the matched classes to detect TPLs.

\subsection{Pre-processing}
{\method} parses the input app and TPL to extract class features and dependency relationships, subsequently constructing their respective CDGs. It then generates a list of candidate app classes for each TPL class based on the similarity of node features between the app and TPL CDGs. By leveraging rich class-level structural information and the resilience of CDGs to obfuscation, {\method} minimizes potential omissions in identifying class correspondences during candidate list generation.

\begin{table}[htbp]
  \caption{Types of CDG edges and features of CDG nodes.}
  \small
  \begin{center}
  \begin{tabular}{clr}
    \toprule
    \textbf{Element} & \textbf{Node/Edge Type} & \textbf{Feature} \\
    \midrule
    \multirow{4}[1]{*}{Node} & static inner class & \textbf{\textit{static}} \\
                              & abstract class & \textbf{\textit{abstract}} \\
                              & other class & \textbf{\textit{default}} \\
                              & interface & \textbf{\textit{interface}} \\

    \midrule
    \multirow{2}[1]{*}{Edge} & extends & \textbf{\textit{extends}} \\
                             & implements & \textbf{\textit{implements}} \\
  \bottomrule
  \end{tabular}
  \label{tab:cdg_type}
  \end{center}
\end{table}

\subsubsection{Class Dependency Graph}\label{subsubsec:cdg}
Class dependency relationships, due to their robustness compared to package structures and their incorporation of semantic and structural information of the code, are utilized by many TPL detection tools~\cite{libpecker,atvhunter}. However, existing tools, in an effort to enhance the feature distinguishability among classes, integrate more granular method and field information within the utilized class dependency relationships, which consequently diminishes their resilience to obfuscation. To address this issue, {\method} constructs CDGs for the app and TPL, wherein classes serve as nodes, as illustrated in Table~\ref{tab:cdg_type}. CDG is a directed graph in which types of edge include \textbf{inheritance} and \textbf{implementation} relationships and types of node include class and interface. Within a complete functional module, these two types of dependencies exhibit resilience against most of obfuscation techniques. The nodes in the CDG are characterized by class modifiers; to enhance resilience against obfuscation, we only consider four types of modifiers: default, abstract, static, and interface. 

\subsubsection{Candidate Class List Generation}\label{subsubsec:class_filter}

Inspired by the message passing concept in graph neural networks, {\method} fully leverages the structural information of CDGs to generate a candidate app class list for each TPL class. As shown in Algorithm~\ref{alg:WL}, {\method} generates features for each node in the CDG (lines 4-9), then calculates the similarity of node features between the app and TPL CDGs (lines 14-17), considering the app class $n_{app}$ as a candidate for the TPL class $n_{lib}$ when the similarity score of the pair $\langle n_{app}, n_{lib} \rangle$ exceeds the threshold $T_c$.
Specifically, {\method} generates a feature for each node $n$ through the following procedure: (1) iteratively aggregating the features of the neighboring nodes of $n$; (2) sorting the aggregated features, concatenating them with the feature of $n$; and (3) applying locality-sensitive hashing (LSH) to generate the node feature. The feature generating process for nodes in the CDGs of the app and TPL iterates $\textbf{\texttt{Diameter}}(G_{lib})$ times. The key \textbf{insight\#1} here is that the feature propagation between the most distant nodes in the TPL's CDG requires exactly as many iterations as the diameter of $G_{lib}$. Fewer iterations may fail to propagate features to all nodes in the CDG, underutilizing the structural information, while more iterations introduce additional computational overhead. This configuration effectively mitigates the impact of graph size differences between the TPL and the app, ensuring a consistent feature generation process and enhancing the reliability of candidate lists generated based on node features.

It is worth noting that {\method} aggregates only the features of the dependent nodes for each node, meaning that the propagation of features occurs in a direction opposite to that of the edges in CDG.
The \textbf{insight\#2} here is that the class dependencies within the app flow from the main program to TPLs. Therefore, the direction of feature propagation described above can mitigate the influence of the main program on feature generation.
{\method} calculates the cosine similarity of features for each node pair \(\langle n_{app}, n_{lib} \rangle\) in the CDGs of app and TPL. Potential candidates are established for pairs with similarity exceeding the threshold $T_c$. The output of preprocessing module of {\method} are the candidate app class lists for all TPL classes \(\mathcal{M}\) and CDGs of the app and TPL.


\begin{algorithm}[b]
\caption{Candidate Class List Generation Algorithm}
\label{alg:WL}

\SetKwFunction{Hash}{LSH}
\SetKwFunction{Sort}{Sort}
\SetKwFunction{Count}{Count}
\SetKwFunction{Equal}{Equal}
\SetKwFunction{Sim}{Sim}
\SetKwFunction{MaxDepth}{\textbf{Diameter}}
\SetKwFunction{FMain}{GenNodesFeature}

\KwIn{$CDG_{app}(V_{app}, E_{app})$, $CDG_{lib}(V_{lib}, E_{lib})$, $T_c$.}
\KwOut{The TPL candidate nodes of app node $\mathcal{M}$.}

\SetKwProg{Fn}{Function}{:}{}
\Fn{\FMain{$G$, $miter$}}{
    $L \leftarrow \{v:\Hash(v) \;|\; \forall v\in V\}$
        
    \For{$iter \leftarrow 1 : miter$}{
        
        \For{each node $u \in V$}{
            $\mathcal{S} \gets \emptyset$ \Comment{Initialized with empty \textbf{multiset}.}
            
            \For{each neighbor $v$ that $(u,v) \in E$}{
                $\mathcal{S} \gets \mathcal{S}\cup L[v]$ \Comment{Add label of $v$.}
            }
            $l \gets L[u] \oplus \Sort(\mathcal{S})$ \Comment{$\oplus$: concatenate.}
            
            $L[u] \gets \Hash(l)$
        }
    }
    
    Return $L$
}

$\mathcal{M} \gets \{\}$

$\mathcal{L}_{app} = \FMain(CDG_{app}, \MaxDepth(CDG_{lib}))$

$\mathcal{L}_{lib} = \FMain(CDG_{lib}, \MaxDepth(CDG_{lib}))$

\For{$c_{l} \in \mathcal{L}_{lib}$}{
    \For{$c_{a} \in \mathcal{L}_{app}$}{
        \If{$\Sim(\mathcal{L}_{lib}[c_{l}], \mathcal{L}_{app}[c_{a}]) > T_c$}{
            $\mathcal{M}[c_{l}] \gets \mathcal{M}[c_{l}]\cup c_{a}$
        }
    }
}

Return $\mathcal{M}$
\end{algorithm}

\subsection{Class Matching}\label{subsec:class_match}

Based on the candidate class lists $\mathcal{M}$, {\method} further conducts fine-grained pairwise matching between app class and TPL class to eliminate false positive candidates.
Class
matching is the process of examining the syntactic and semantic consistency between each candidate app class $c_a$ and TPL class $c_l$ in $\mathcal{M}$. 
{\method} categorizes classes into \textit{stateful classes} and \textit{stateless classes} based on the presence of non-static fields. Stateful classes contain non-static fields, allowing changes in field values to directly reflect the object's state; conversely, stateless classes lack fields, making state changes less perceptible. Based on this distinction, {\method} processes stateful and stateless classes separately, leveraging code syntactics and semantics for effective class matching.
{\method} first verifies whether the pair $\langle c_a, c_l\rangle$ exhibits consistent categories, terminating the matching process if the categories differ.

\noindent \textbf{Slice-based Member Matching.}
To mitigate the impact of fine-grained code obfuscation (e.g., control flow flattening), {\method} performs taint analysis using the method parameters as the source and the method exit as the sink. {\method} slices the instructions based on data flows from source to sink, extracting the opcodes of the sliced instructions as the method's functional representation. For methods without parameters, {\method} extracts the opcodes of all instructions. For class pair $\langle c_a, c_l\rangle$, {\method} calculates the opcodes overlap rate between the method $m_{a} \in c_{a}$ in the app and the method $m_{l} \in c_{l}$ in the TPL using the following formula:
\begin{align}
    overlap_{m_a, m_l} = \frac{|Op(m_{a}) \cap Op(m_l)|}{|Op(m_l)|},
\end{align}
where $Op(\cdot)$ denotes the set of opcodes associated with the input method, and $|\cdot|$ represents the cardinality of the set.
{\method} extracts all method pairs that satisfy $Overlap_{m_a,m_l}>T$ and share the same method fuzzy signature~\cite{libscan}, prioritizing the establishment of correspondences between pairs with the highest overlap rate, while ensuring that each method is matched at most once.
Note that this process does not guarantee the successful matching of all methods in $c_a$ and $c_l$. In the presence of method additions or deletions, which may result from obfuscations, the correspondence may diminish. {\method} approach disregards the unmatched methods, thereby focusing on the matched methods and mitigating the impact of such obfuscations.

For stateless pairs, {\method} calculates the proportion $R_m$ of matched methods among all methods in $c_l$, and the proportion $R_o$ of opcodes of matched methods to the total opcodes in $c_l$ to determine the confidence score of matching:
\begin{align}
    CMS{c_a, c_l} = \frac{R_m+R_o}{2},
\end{align}
Class $c_a$ and class $c_l$ are considered matched when $CMS(c_a, c_l) > T$.
The class matching confidence is determined by integrating both ratios, thus mitigating the uneven distribution of opcodes within methods due to obfuscation. For instance, if a obfuscator inserts a string decryption method in a simple app class, which constitutes a large proportion of the opcodes in the class, but no matching method exists in the corresponding TPL class, considering opcode matching alone could result in a false negative.


Stateful classes can be more complex, and similar functionalities or operation patterns may lead to similar opcode sets, but the fields involved may differ, resulting in cases of similar syntax but different semantics. To further eliminate false positives, {\method} matches fields using field read and write operations from the matched method pairs, provided that $CMS(c_a, c_l) > T$. Based on the type and frequency of field-related operations, {\method} performs field matching. If the match fails, it indicates a false positive in method matching, and the method match is discarded. This step allows {\method} to leverage field information to eliminate method matching false positives while enabling field matching, thereby providing a foundation for deeper semantic analysis.

\noindent \textbf{Functionality Summary-based Class Matching.}
Existing research~\cite{libscan} utilize the opcodes of call chains to further mitigate false positives in method matching. However, the extensive code introduced by call chains may weaken the distinction between the opcode sets of app method and TPL method, and the generation and traversal of call graphs incur substantial runtime overhead. In contrast, {\method} eliminates false positives in class matching by generating method call sequences within classes to construct diverse contexts for semantic consistency verification. Subsequently, {\method} traverses these sequences to generate summaries representing class functionality, using similarity measures to assess the semantic consistency between the two classes $\langle c_a, c_l\rangle$.
The method call sequence $CS_c=\{m_1,m_2,...,m_n\}$ of a class $c$ is an ordered list composed of $n$ methods within the class. {\method} simulates actual usage scenarios to generate method call sequences $CS_{c_l}$ of TPL class $c_l$. Initially, {\method} simulates object instantiation by randomly selecting a constructor \texttt{<init>} of $c_l$ to add to the sequence. Furthermore, {\method} randomly selects methods in the $c_l$, excluding constructors, to include in the call sequence, repeating this process $S$ times to simulate object usage. Based on the matched methods, {\method} constructs the method call sequence $CS_{c_a}$ for the corresponding app class $c_a$.

To mitigate the impact of noise introduced by obfuscation, {\method} distills code of each method call sequence into field operations that reflects the state changes of the object created and used through the method call sequence. Specifically, {\method} identifies three types of field operations through static code analysis: \textit{initialization}, \textit{assignment}, and \textit{method invocation}. 
For each field operation, {\method} extracts used key elements, including the names and fuzzy types of matched fields, as well as the positions and fuzzy types of method parameters. Therefore, each method call sequence $CS$ forms a corresponding field operation sequence $FOS$.

Since classes retain the same semantics despite obfuscation, the field operation sequence $FOS$ should also exhibit similarity before and after obfuscation (\textbf{insight\#3}). 
Therefore, {\method} first replaces field names in the field operation sequence with unique number to eliminate the impact of identifier obfuscation, then computes locality sensitive hashing of field operations to generate class functional summaries $FS$. To address potential disruptions in field operation order caused by control flow obfuscation, {\method} applies the Hungarian algorithm~\cite{Hungarian_algorithm} ($\mathcal{H}(\cdot)$) to maximize matches between the summaries of app class $FS_{c_a}$ and TPL class $FS_{c_l}$. The proportion of these maximum matches $|\mathcal{H}(FS_{c_a}, FS_{c_l})|$ to the total number of summaries $|FS_{c_l}|$ is used as the semantic similarity score for the method call sequences $CS_{c_a}$ and $CS_{c_l}$.
To avoid false positives caused by random method selection, {\method} generates $K$ method call sequences to construct diverse code context, and the average of the semantic matching scores of all call sequences is used as the class matching confidence score $CMS$.
\begin{align}
    CMS_{c_a,c_l}=\frac{1}{K}\sum_{i=1}^{K}\frac{|\mathcal{H}(FS_{c_a}^i,FS_{c_l}^i)|}{|FS_{c_l}^i|}
\end{align}

{\method} regards pairs with matching confidence scores exceeding $T$ as matches. Specifically, nodes with confidence score $CMS_{c_a,c_l}$ greater than $T_h$ ($T_h > T$) are identified as \textbf{high-confidence matches} and are excluded from subsequent matching processes.


\subsection{TPL Detection}\label{subsec:tpld}
To capture the differences between different versions of TPLs, {\method} performs an analysis across all versions of the given TPL. Since the class matching module considers only intra-class information, it can incorrectly yield high matching confidence scores for structurally dissimilar classes in scenarios such as code reuse or code cloning. To address this issue, {\method} leverages the structural similarity between CDGs to verify whether matched classes remain structurally consistent. However, identifying a small graph (TPL) within a larger graph (app) is fundamentally a subgraph isomorphism problem, which is one of NP problems and computationally expensive~\cite{ullmann2011bit,vf2}, and obfuscation may compromise isomorphism. 

We observe that when an app utilizes the TPL class $c_l$, to ensure functional integrity, the classes on which $c_l$ depends must also be incorporated into the app. This dependency relationship recursively extends until encountering dependency-free classes (i.e., terminal nodes with an out-degree of 0 in the CDG), naturally forming a functional module of the TPL. 
Therefore, to eliminate potential false positives, {\method} inspects whether the matched classes of TPL form at least one path $P_{lib}$ to the terminal nodes in the CDG. If $P_{lib}$ exists, {\method} then examines whether the corresponding set of app nodes $N_{app}^{P_{lib}}$ also forms an identical path $P_{app}$ in the app's CDG, where an identical path denotes passing through edges with identical features in the same order. If \( P_{lib} \) and \( P_{app} \) exist and \( P_{lib} = P_{app} \), {\method} considers that the CDGs of the app and TPL have similar structures.


After preliminarily verifying structural similarity, {\method} calculates a confidence score, $Score_{app,lib}$, to assess the likelihood that a given TPL is integrated in the app, based on the class matching results between the app and the TPL. 
For the sake of clarity in notation, we define the set of high-confidence matched pairs as $HM_{app,lib} = \{(c_a, c_l) |T_h\leq CMS_{c_a, c_l} \}$, and the set of other matched pairs as $M_{app,lib} = \{(c_a, c_l) | T \leq CMS_{c_a, c_l} < T_h\}$. {\method} uses the following formula to calculate the confidence score for TPL detection:
\begin{align}
    Score_{app,lib}=\frac{|M_{app,lib}|+\alpha\times|HM_{app,lib}|}{|\mathcal{M}|}
\end{align}
where $|\mathcal{M}|$ denotes the number of TPL classes annotated with candidate app classes (\S\ref{subsubsec:class_filter}). 
$\alpha$ represents the weight of high-confidence matches, and we set it to 1.5.
When \( Score_{app,lib} > T_G \), {\method} considers that \( lib \) is used by \( app \).
Since slight differences among versions may result in scores exceeding $T_G$, {\method} prioritizes selecting the TPL version with the highest score as the final result. If multiple versions share the same highest score, {\method} further selects the version with the largest number of high-confidence matched classes in the TPL's CDG as the final result. 
In certain extreme cases, the differences between various versions of TPL may be minimal (for instance, merely a change in a version-identifying string). Therefore, {\method} extracts a set of literal to construct feature set $\mathbb{I}_{lib_{v_i}}$ from the matched classes of different versions $lib_{v_i}$, and then extracts a set of literal features $\mathbb{I}_{app}$ from the matched classes of the app. Finally, {\method} calculates the intersection of the literal features of the TPL version $lib_{v_i}$ and the app, and selecting the version with the largest intersection as the final version.



\section{Experiments}
In this section, we conduct experiments which are designed to answer the following four research questions.

\begin{itemize}[leftmargin=*]
    \item \textbf{RQ1 (Effectiveness)}: Can {\method} achieve a higher F1 score compared to state-of-the-art TPL detection tools?
    
    \item \textbf{RQ2 (Reliability)}: Does {\method} outperform existing TPL detection tools in class-level detection?

    \item \textbf{RQ3 (Contribution of Components)}: Do the main components of {\method} contribute significantly to its performance?

    \item \textbf{RQ4 (Efficiency)}: How does {\method}'s efficiency compare with state-of-the-art TPL detection tools?
\end{itemize}

\subsection{Experimental Setup}\label{subsec:metric}
\noindent \textbf{Datasets.} We utilize the dataset constructed by LibScan~\cite{libscan} to evaluate the performance of various TPL detection tools. This dataset comprises 1,231 apps (\#app and \#Tuning) and 562 TPL versions as the TPL database, as presented in Table~\ref{tab:dataset}. To better compare the effectiveness of TPL detection tools across different types of apps, we divide it into three subsets: the unobfuscated dataset $D_1$, the obfuscated dataset $D_2$, and the optimized dataset $D_3$. 
The obfuscated dataset is generated using three obfuscation tools—Allatori, DashO, and ProGuard—with different configurations. Specifically, DashO employs four distinct obfuscation levels to produce corresponding obfuscated apps, including control flow randomization (cfr), package flattening and identifier renaming (pf-ir), dead code removal (dcr), and a combination of the three obfuscations (cfr-pf-ir-dcr).
Furthermore, we identified errors in the ground truth of the LibScan dataset. After manual review and correction, the total ground truth for $D_1$ and $D_2$ increased from 5,956 to 6,168.
The optimized dataset $D_3$ consists of 51 apps compiled with D8 and 51$\times$3 apps compiled with R8 under different optimization configurations. R8-shrink-opt enables optimization on top of code shrinking, while R8-shrink-orlis applies Orlis’s ProGuard configuration to perform repackaging and renaming obfuscation in addition to code shrinking.
The 453 TPL versions in $D_1$ and $D_2$ encompass 236 distinct TPLs, with each TPL averaging $\sim$2 versions. In $D_3$, the 109 TPL versions correspond to 59 unique TPLs, demonstrating a version-to-TPL ratio of 1.85 versions per TPL.

\begin{table}[htbp]
  \caption{Statistic of dataset used to evaluate {\method}.}
  \small
  \begin{center}
  \begin{tabular}{ccccc}
    \toprule
    \textbf{Dataset} & \textbf{Category} & \textbf{\#apps} & \textbf{\#Tuning} & \textbf{\#libs} \\
    \hline
    
    $D_1$ & Non-obfus & 203 & 22 & \multirow{7}[1]{*}{453} \\
    \cline{1-4}
    
    \multirow{6}[1]{*}{$D_2$} & Allatori & 188 & 22 &  \\
    \cline{2-4}
    & DashO-cfr & 79 & 9 & \\
    \cline{2-4}
    & DashO-pf-ir & 79 & 9 & \\
    \cline{2-4}
    & DashO-dcr & 79 & 9 & \\
    \cline{2-4}
    & DashO-cfr-pf-ir-dcr & 159 & 22 & \\
    \cline{2-4}
    & ProGuard & 152 & 17 & \\

    \hline
    \multirow{4}[1]{*}{$D_3$} & D8-compiled & 46 & 5 & \multirow{4}[1]{*}{109} \\
    \cline{2-4}
    & R8-shrink & 46 & 5 & \\
    \cline{2-4}
    & R8-shrink-opt & 46 & 5 & \\
    \cline{2-4}
    & R8-shrink-orlis & 46 & 5 & \\
    \bottomrule
    
  \end{tabular}
  \label{tab:dataset}
  \end{center}
\end{table}

\newcolumntype{M}[1]{>{\centering\arraybackslash}m{#1}}   

\begin{table*}[!ht]
  \caption{Detection performance (\%) of TPL detection tools ({\method}, LibPecker, LibHunter, LibScan, LIBLOOM) on datasets $D_{1}$ and $D_{2}$ at library-level, version-level, and version-level$^{\dagger}$.}
  \begin{center}
  \small
  \begin{tabular}{clM{1.1cm}M{1.1cm}M{1.1cm}M{1.1cm}M{1.1cm}M{1.1cm}M{1.1cm}M{1.1cm}M{1.1cm}}
    \toprule
    \multirow{2}[1]{*}{\textbf{Dataset}} & \multirow{2}[1]{*}{\textbf{Tools}} & \multicolumn{3}{c}{\textbf{Library-level}} & \multicolumn{3}{c}{\textbf{Version-level}} & \multicolumn{3}{c}{\textbf{Version-level$^{\dagger}$}} \\

    \cmidrule[0.5pt](lr){3-5} \cmidrule[0.5pt](lr){6-8} \cmidrule[0.5pt](lr){9-11}
    & & Precision & Recall & F1 & Precision & Recall & F1 & Precision & Recall & F1 \\

    \cmidrule[0.5pt]{1-11}
    \multirow{5}[1]{*}{$D_1$} & LibPecker & 80.86 & \textbf{99.54} & 89.23 & 78.82 & 97.03 & 86.98 & 64.73 & 97.03 & 77.65 \\

    & LibHunter &  80.26 & 93.93 & 86.56  & 80.03 & 93.66 & 86.31  & 61.96 & 93.66 & 74.59 \\
    
    & LibScan &  96.96 & 98.88 & 97.91 & 96.83 & \textbf{98.75} & 97.78 & 68.50 & \textbf{98.75} & 80.89  \\
    
    & LIBLOOM & \textbf{99.00} & 98.28 & \textbf{98.64} & 95.74 & 95.05 & 95.40 & 67.92 & 95.05 & 79.23 \\
    
    & {\method} & \underline{97.91} & \underline{99.14} & \underline{98.52} & \textbf{97.26} & \underline{98.48} & \textbf{97.87} & \textbf{79.03} & \underline{98.48} & \textbf{87.69} \\

    \cmidrule[0.5pt]{1-11}

    \multirow{5}[1]{*}{$D_2$} & LibPecker & 73.41 & 71.49 & 72.44 & 66.89 & 65.14 & 66.01 & 56.30 & 65.14 & 60.40 \\

    & LibHunter & 80.41 & 59.89 & 68.65 & 79.00 & 58.84 & 67.45 &  61.81 & 58.84 & 60.29 \\
    
    & LibScan & 98.03 & 95.02 & 96.50 & \textbf{96.50} & 93.54 & 95.00 & 64.05 & 93.54 & 76.04 \\
    
    & LIBLOOM & \textbf{98.74} & 92.58 & 95.56 & 95.86 & 89.88 & 92.77 & 66.49 & 89.88 & 76.43 \\							
    & {\method} & 96.66 & \textbf{98.65} & \textbf{97.64} & 94.39 & \textbf{96.33} & \textbf{95.35} & \textbf{75.76} & \textbf{96.33} & \textbf{84.82} \\

    \bottomrule
    \multicolumn{10}{l}{\textbf{bold}: the highest value, \underline{underlined}: the second-ranked value.}
  \end{tabular}
  \label{tab:perf_all}
  \end{center}
\end{table*}

\noindent \textbf{Metrics.}
We evaluate the performance of TPL detection tools by calculating their F1 score. To quantify library-level detection performance, we adopt the same approach as LibScan to count the \textit{true positives} ($TP_l$), \textit{false positives} ($FP_l$), and \textit{false negatives} ($FN_l$). For version-level detection, LibScan counts $TP_v$, $FP_v$ and $FN_v$, ensuring that $TP_l + FP_l = TP_v + FP_v$. 
This reflects LibScan's relaxed false positive statistical method at the version-level. For instance, if a tool reports \texttt{gson-2.4} and \texttt{gson-2.5} in an app, while the actual version is \texttt{gson-2.6}, LibScan records this as both a false negative and a false positive at the version-level. In a similar case, if a tool detects \texttt{gson-2.4}, \texttt{gson-2.5}, and \texttt{gson-2.6} in an app, LibScan considers it a true positive and disregards the incorrect versions. We argue that this statistical approach introduces bias in the assessment of version-level detection performance for TPL detection tools, as a tool could simply report all \texttt{gson} versions to easily achieve a false positive rate (FPR) of 0 at the version-level.

To more accurately assess the version-level detection performance of TPL detection tools, we propose a new method for counting true positives ($TP_{v^{\dagger}}$), false positives ($FP_{v^{\dagger}}$), and false negatives ($FN_{v^{\dagger}}$) at the version-level, referred to as \textbf{version-level$^{\dagger}$} to distinguish it from LibScan's approach. Specifically, if a tool correctly identifies the actual version among reported versions of the \texttt{gson} TPL, we count it as one $TP_{v^{\dagger}}$, while the remaining incorrect versions are counted as $FP_{v^{\dagger}}$. If the correct version is not included in the reported list, we count one $FN_{v^{\dagger}}$ and classify all reported versions as $FP_{v^{\dagger}}$. Consequently, version-level$^{\dagger}$ detection satisfies the condition $TP_{v^{\dagger}} + FP_{v^{\dagger}} \geq TP_v + FP_v$.

\subsection{Thresholds Tuning}
To avoid biases introduced by empirical threshold settings, we extract 10\% of apps from different datasets to construct validation datasets (shown in Table~\ref{tab:dataset} \#Tuning) for thresholds tuning of {\method} and the baseline tools (i.e. LibScan, LIBLOOM, LibHunter, and LibPecker). 
Following LibScan's approach~\cite{libscan}, we perform thresholds tuning separately on the obfuscated datasets $D_1$, $D_2$ and the optimized dataset $D_3$.

\setlength{\tabcolsep}{3pt}
\begin{table}[!htbp]
  \caption{Thresholds tuning results of different TPL detection tools.}
  \small
  \begin{center}
  \begin{tabular}{lllll}
    \toprule
    \textbf{Tools} & \multicolumn{2}{c}{\textbf{Obfuscation}} & \multicolumn{2}{c}{\textbf{Optimization}} \\
    
    \cmidrule[0.5pt]{1-5}
    
    LibPecker & $T_{lib}=0.7$ & $T_{pkg}=0.5$ & $T_{lib}=0.1$ & $T_{pkg}=0.05$  \\
    LibHunter & $T_{mtd}=0.3$& $T_{lib}=0.95$ & $T_{mtd}=0.9$ &$T_{lib}=0.2$ \\
    LibScan & $T_{class}=0.7$& $T_{lib}=0.85$ & $T_{class}=0.7$& $T_{lib}=0.1$ \\
    LIBLOOM & $T_{pkg}=0.05$& $T_{lib}=0.9$& $T_{pkg}=0.95$& $T_{lib}=0.15$ \\
    {\method} & $T=0.7$ & $T_G=0.85$  & $T=0.8$ & $T_G=0.5$ \\

    \bottomrule
    
  \end{tabular}
  \label{tab:tune}
  \end{center}
\end{table}

For the threshold $T_c$ of the candidate list construction, we initially constructs the ground truth of class correspondences between the apps and TPLs in the validation set, after which we employs the candidate class list construction algorithm to check whether the candidates of TPL classes contains the corresponding app classes to compute \textit{false negative rate} (FNR) and \textit{false positive rate} (FPR). Starting from 0, $T_c$ is gradually increased to 1.0 in steps of 0.05. Eventually, a threshold of 0.85 is set to balance the FNR and FPR. The threshold $T_h$ indicates high-confidence matches, which is fixed at 0.9, invariant to the dataset. For the class matching threshold $T$ and the CDG matching threshold $T_G$, we perform thresholds tuning on the validation set using grid search, with each threshold ranging from 0 to 1.0, incremented by steps of 0.05. Subsequently, the thresholds yielding the highest F1 score are selected for subsequent experiments. 
Since the baseline tools also involve two thresholds, we apply the same grid search procedure to choose the best thresholds. 
The thresholds tuning results are presented in Table~\ref{tab:tune}.

\subsection{RQ1: Effectiveness of {\method}}\label{subsec:effectiveness}
We utilized $D_1$, $D_2$ and $D_3$ to evaluate the performance of {\method} in detecting TPLs and their versions in unobfuscated apps, obfuscated apps, and optimized apps. The evaluations are conducted at the library-level, version-level, and version-level$^{\dagger}$, and results are compared against four state-of-the-art TPL detection tools: LibScan~\cite{libscan}, LIBLOOM~\cite{libloom}, LibHunter~\cite{libhunter}, and LibPecker~\cite{libpecker}.
Table~\ref{tab:perf_all} presents the detailed detection performance of {\method} and baseline tools on datasets $D_1$ and $D_2$. Due to the stricter statistical method of false positives (\S\ref{subsec:metric}), all tools exhibit a notable performance decline at the version-level$^{\dagger}$. We find that {\method} achieves detection performance comparable to state-of-the-art TPL detection tools at both the library- and version-level, while significantly outperforming the baselines at the version-level$^{\dagger}$. 
Specifically, {\method} achieves an average improvement in F1 scores at the version-level$^{\dagger}$ of 12.39\% and 25.91\% on the unobfuscated dataset $D_1$ and obfuscated dataset $D_2$, respectively, reaching 87.69\% and 84.82\%. This improvement is primarily attributed to {\method}'s ability to significantly reduce the number of false positive versions while maintaining a low false negative rate, demonstrating its effectiveness.
As described in \S\ref{subsec:metric}, the relaxed version-level FP evaluation inflates the performance of detection tools, making it appear comparable to library-level performance. However, a comparison between version-level and version-level$^{\dagger}$ FPs reveals that advanced tools such as LibScan and LIBLOOM report a substantial number of FP versions, leading to significant performance discrepancies. In contrast, {\method} achieves FPs at only 58.70\% and 68.04\% of those reported by LibScan and LIBLOOM, respectively, while exhibiting the low false negative rate, resulting in superior performance at version-level$^{\dagger}$. 



\setlength{\tabcolsep}{2.8pt}
\begin{table}[!htbp]
  \caption{The F1 score (\%) of TPL detection tools on different obfuscation level ($D_{1}$ and $D_2$).}
  \begin{center}
  \small
  \begin{tabular}{clccccc}
    \toprule
    \textbf{Level} & \textbf{Type} & LibPecker & LibHunter & LibScan & LIBLOOM & {\method} \\
    \midrule
    \multirow{7}[1]{*}{\textbf{Lib}}& Non-obfus & 89.23 & 86.56 & 97.91 & \textbf{98.64} & \underline{98.52} \\
    & Allatori & 75.26 & 44.05 & 94.78 & 93.33 & \textbf{96.74} \\
    & DashO-cfr & 76.67 & 85.19 & 98.27 & \textbf{99.43} & \underline{98.97} \\
    & DashO-pf-ir & 71.90 & 85.90 & 98.27 & 93.56 & \textbf{98.98} \\
    & DashO-dcr & 71.69 & 84.64 & 92.53 & 95.09 & \textbf{95.17} \\
    & DashO-A & 49.04 & 44.43 & 97.12 & 94.69 & \textbf{97.24} \\
    & ProGuard & 87.64 & 88.38 & 98.49 & \textbf{99.16} & 98.45 \\

    \midrule

    \multirow{7}[1]{*}{\textbf{Ver$^{\dagger}$}}& Non-obfus & 77.65 & 74.59 & 80.89 & 79.23 & \textbf{87.63} \\
    & Allatori & 60.78 & 38.39 & 75.62 & 72.76 & \textbf{84.24} \\
    & DashO-cfr & 62.08 & 73.08 & 76.95 & 80.72 & \textbf{86.35} \\
    & DashO-pf-ir & 57.56 & 73.78 & 76.31 & 75.73 & \textbf{87.11} \\
    & DashO-dcr & 57.32 & 72.44 & 58.40 & 76.91 & \textbf{85.22} \\
    & DashO-A & 43.69 & 43.56 & 75.16 & 76.53 & \textbf{82.09} \\
    & ProGuard & 75.64 & 75.55 & 80.27 & 79.89 & \textbf{86.85} \\

    \bottomrule
    \multicolumn{7}{l}{Lib: Library-level, Ver$^{\dagger}$: Version-level$^{\dagger}$, DashO-A: DashO-cfr-pf-ir-dcr.}
  \end{tabular}
  \label{tab:perf_obfs}
  \end{center}
\end{table}

\noindent \textbf{Resilience to Obfuscations.}
Comparing the detection performance of different TPL detection tools on the obfuscated dataset $D_2$ in Table~\ref{tab:perf_all}, we find that LibPecker reports the highest FPs, while LibHunter reports the most FNs, resulting in poorer F1 score at library-level and version-level$^{\dagger}$ for both tools. The primary reason is that LibPecker uses coarse-grained class dependency signatures for matching, which fail to effectively distinguish between classes from different TPLs under obfuscation, leading to numerous FPs. In contrast, LibHunter's fine-grained features depend on  strings, which cannot be successfully matched under string encryption obfuscation, resulting in a high number of FNs. In comparison, {\method}, LibScan, and LIBLOOM exhibit higher resistance to obfuscation. 
{\method} achieves the highest F1 score at the library-level, version-level, and version-level$^{\dagger}$. Notably, by effectively capturing fine-grained differences across versions to eliminate FPs, {\method} improves the F1 score at version-level$^{\dagger}$ by 10.98\% over the best existing tool, LIBLOOM, reaching 84.82\%, making it the most effective tool. Furthermore, LibScan and LIBLOOM exhibit approximately 15-fold and 11-fold increases in FPs at version-level$^{\dagger}$ compared to version-level, which are obscured by the more relaxed FP statistical criteria of version-level. Consequently, their performance deteriorates most significantly at version-level$^{\dagger}$, indicating that obfuscation has a substantial impact on their version identification capabilities.



Under the DashO-cfr configuration, LIBLOOM exhibits superior F1 performance compared to {\method} at the library-level. However, at the version-level$^{\dagger}$, it underperforms relative to {\method}. This indicates that while coarse-grained features utilized by LIBLOOM are unaffected by control flow randomization obfuscation, they struggle to distinguish TPL versions. In contrast, {\method} achieves a more effective balance between obfuscation resilience and version identification performance.
Table~\ref{tab:perf_obfs} presents the F1 scores of TPL detection tools under different obfuscation levels, highlighting the varying resilience of these tools to different obfuscation configurations. Overall, Allatori and DashO-A are the most impactful obfuscation configurations, resulting in the lowest F1 scores for LibPecker, LibHunter, LIBLOOM, and {\method}. In contrast, LibScan achieves the highest F1 score under DashO-A and the lowest under DashO-dcr. This discrepancy is due to the significant impact of dead code elimination on the opcode features utilized by LibScan, while the DashO-A configuration applies weaker dead code elimination on apps. Notably, {\method} consistently achieves significantly higher F1 scores than baseline tools across all obfuscation configurations at the version-level$^{\dagger}$, demonstrating {\method}'s resilience against obfuscation in version identification.

\setlength{\tabcolsep}{2pt}
\begin{table}[htbp]
  \caption{The F1 score (\%) of TPL detection tools on different optimization level ($D_{3}$).}
  \begin{center}
  \small
  \begin{tabular}{clccccc}
    \toprule
    \textbf{Level} & \textbf{Type} & LibPecker & LibHunter & LibScan & LIBLOOM & {\method} \\
    \midrule
    \multirow{4}[1]{*}{\textbf{Lib}}& D8 & 71.79 & 65.77 & 75.28 & \textbf{84.07} & 71.58 \\
    & shrink & 77.09 & 74.79 & 69.06 & \textbf{77.82} & 69.18 \\
    & shrink-opt & \textbf{77.42} & 57.76 & 15.14 & 61.48 & 47.16 \\
    & shrink-orlis & 70.10 & \textbf{75.21} & 68.59 & 72.79 & 68.94 \\

    \cline{2-7}
    & Overall & 73.78 & 68.84 & 62.83 & \textbf{75.16} & 66.04$^{\downarrow 9.12}$ \\

    \midrule

    \multirow{4}[1]{*}{\textbf{Ver$^{\dagger}$}}& D8 & 63.71 & 61.40 & 23.17 & \textbf{71.49} & 59.96 \\
    & shrink & 49.22 & \textbf{67.57} & 36.71 & 54.82 & 46.71 \\
    & shrink-opt & 40.13 & 35.21 & 13.53 & \textbf{44.85} & 28.57 \\
    & shrink-orlis & 42.68 & \textbf{67.93} & 37.04 & 52.10 & 51.83 \\

    \cline{2-7}
    & Overall & 50.10 & \textbf{59.54} & 27.73 & 57.54 & 49.31$^{\downarrow 10.23}$ \\

    \bottomrule
    \multicolumn{7}{l}{Lib: Library-level, Ver$^{\dagger}$: Version-level$^{\dagger}$.}
  \end{tabular}
  \label{tab:perf_optim}
  \end{center}
\end{table}


    
    
    

\noindent \textbf{Resilience to Optimizations.}
Although {\method} is primarily designed for obfuscated apps, a certain proportion of real-world apps undergo optimization~\cite{libhunter}. Therefore, we further evaluate the resilience of {\method} against optimization using the optimized dataset $D_3$ constructed by~\cite{libscan}. Table~\ref{tab:perf_optim} presents the F1 scores of various TPL detection tools under different optimization levels. 
LibHunter, designed for optimized apps, achieves the highest F1 score at the version-level$^{\dagger}$, whereas LIBLOOM performs best at the library-level, reflecting its high resilience to optimization due to the use of coarse-grained features for library-level detection. Surprisingly, LibScan performs the worst at both the library- and version-level$^{\dagger}$, primarily due to its reliance on fine-grained opcode features, which fail to handle optimization techniques, particularly under the shrink-opt configuration.

Overall, {\method} experiences a decline in F1 scores, with decreases of 9.12 and 10.23 compared to the best-performing tools at the library- and version-level$^{\dagger}$, respectively. Further analysis reveals that the relatively low F1 score of {\method} is mainly attributed to a high number of FNs. Manual inspection identifies two primary causes: \ding{172} incorrect TPL versions in the ground truth. Since the correct versions are not included in the TPL database and cannot be accessed, {\method}'s class matching module fails due to version discrepancies, ultimately affecting TPL detection. \ding{173} Excessive method reduction and removal, leading to a high number of method matching failures, which similarly impact TPL detection performance.

\subsection{RQ2: Class-level Performance}\label{subsec:classlevel}
Although {\method} demonstrates high performance at the library-level and version-level$^{\dagger}$ (\S\ref{subsec:effectiveness}) on obfuscated apps, some downstream tasks, such as TPL removal and isolation~\cite{orlis}, require fine-grained identification of TPL code, thus imposing demands on class-level detection performance of TPL detection tools. 
Moreover, due to the complexity of the obfuscation and optimization, the actual range of codes relied upon by TPL detection tools when reporting TPLs remains unknown, leading to a lack of understanding regarding the reliability of tools. 
Therefore, we establish a one-to-one mapping between each class in the TPL and those in the app to construct the ground truth for evaluating class-level detection performance of TPL detection tools. Since the class mapping files of apps are unavailable, we manually analysis the decompiled code of both the apps and TPLs to construct the ground truth. Given the substantial workload of manual analysis, we randomly selected 20 pairs of $\langle app, lib \rangle$ from the intersection of the TPs reported by different tools in datasets $D_1$, $D_2$, and $D_3$ for evaluation.
As LibPecker does not explicitly establish mapping between classes, we do not evaluate its class-level detection performance.

\begin{table}[htbp]
  \caption{Class-level detection performance of different TPL detection tools.}
  \small
  \begin{center}
  \begin{tabular}{lccccc}
    \toprule
     \textbf{Metric} & LibPecker & LibHunter & LIBLOOM & LibScan & {\method}\\
    \cmidrule[0.5pt]{1-6}

    \textbf{Precision} & - & 93.06 & 91.67 & \textbf{98.59}& \underline{96.70}  \\
    \textbf{Recall} & - & 69.79 & 68.75 & 72.92 & \textbf{91.67} \\
    \textbf{F1} & - & 79.76 & 78.57 & 83.83 & \textbf{94.12}\\

    \bottomrule
    
  \end{tabular}
  \label{tab:classlevel}
  \end{center}
\end{table}

As shown in Table~\ref{tab:classlevel}, LIBLOOM, which performs well in detection on obfuscated and optimized datasets, exhibits the poorest class-level F1. We find that LIBLOOM's method of hashing extracted features and using bloom filters for TPL detection inevitably leads to hash collisions, resulting in more false positives and false negatives at the class-level, such as incorrectly matching TPL classes with app interfaces. 
The class-level detection performance reflects the contribution of TPL classes to the tool's library- and version-level$^{\dagger}$ detection results. A TPL detection tool with a low class-level F1 score is likely to report unreliable library- and version-level$^{\dagger}$ TPs, as the identified TPL classes are not actually part of the TPL. This may lead to unpredictable detection results for different apps using the same TPL.
{\method}'s class-level F1 score outperforms baseline tools, reflecting its superior reliability. The reason lies in {\method}'s fine-grained handling of stateful classes, which eliminates syntactically similar but semantically distinct candidate app classes by leveraging field-related operations and class functionality summaries (\S\ref{subsec:class_match}).
Moreover, the class-level recall of {\method} significantly outperforms the baseline tool, enabling it to better support downstream tasks that rely on specific TPL code.


\subsection{RQ3: Contribution of Components}\label{subsec:ablation}
To evaluate the effectiveness of {\method} in generating candidate lists for TPL classes by using structural information of CDGs, we replaced the candidate app class list generation algorithm with a signature matching algorithm to implement {\method}-f. This algorithm generates class signature set for each class by extracting fuzzy signatures of its members~\cite{libscan}, then generating a candidate app class list for the TPL class based on the overlap rate of the signature set. Furthermore, since many existing tools focus on opcodes for class matching, we aim to assess whether the class summary-based class matching step in {\method} significantly improves detection performance, determining its necessity. Therefore, we perform method matching based on the opcodes of methods and verify the reliability of method matches by checking field read and write operations to eliminate false positives. We then compute the ratio of opcodes in matched methods to the total opcodes of the TPL class as the class matching confidence score. This baseline removes the step of the {\method} class matching module that generate method call sequences and class functionality summaries for semantic matching, which is denoted as {\method}-s.

As shown in Table~\ref{tab:abstudy}, {\method} markedly outperforms two baselines, achieving an average F1 score improvement of 22.86\% and 25.32\% at library-level and version-level$^{\dagger}$ on three datasets, respectively, thereby demonstrating the effectiveness of the candidate class list generation algorithm and semantic matching step in class matching. By leveraging the structural information of the CDGs, {\method} identifies candidate app classes for each TPL class, reducing the overhead of fine-grained class matching while effectively capturing structurally similar classes. In contrast, {\method}-f, which adopts a more relaxed method and suffers from higher false positive rates. {\method}-s, after removing the semantic matching step, fails to distinguish between different TPLs due to its inability to exploit semantic-level information, leading to a notable decline in both library- and version-level$^{\dagger}$ detection performance.

\setlength{\tabcolsep}{4pt}
\begin{table}[htbp]
  \caption{Effectiveness of {\method} and baselines on $D_1$, $D_2$ and $D_3$.}
  \small
  \begin{center}
  \begin{tabular}{clcccccc}
    \toprule
     \multirow{2}[1]{*}{\textbf{Dataset}} & \multirow{2}[1]{*}{\textbf{Tools}} & \multicolumn{3}{c}{\textbf{Library-level}} &  \multicolumn{3}{c}{\textbf{Version-level$^{\dagger}$}} \\
    \cmidrule[0.5pt](lr){3-5} \cmidrule[0.5pt](lr){6-8}
    & & P & R & F1 & P & R & F1 \\
    
    \cline{1-8}
    
    \multirow{3}[1]{*}{$D_1$} & {\method}-f & 84.00 & 96.91 & 90.00 & 60.82 & 80.34 & 69.23 \\

    & {\method}-s & 75.89 & 90.40 & 82.51 & 53.32 & 75.50 & 62.50  \\
    
    & {\method} & \textbf{97.85} & \textbf{99.08} & \textbf{98.46} & \textbf{78.97} & \textbf{98.42} & \textbf{87.63} \\

    \cline{1-8}

    \multirow{3}[1]{*}{$D_2$} & {\method}-f & 85.63 & 75.74 & 80.38 & 59.10 & 59.41 & 59.25 \\

    & {\method}-s & 77.29 & 66.00 & 71.20 & 46.69 & 50.05 & 48.31 \\
    
    & {\method} & \textbf{96.66} & \textbf{98.65} & \textbf{97.64} & \textbf{75.76} & \textbf{96.33} & \textbf{84.82} \\

    \cline{1-8}

    \multirow{3}[1]{*}{$D_3$} & {\method}-f & 64.72 & 30.85 & 41.78 & 47.94 & 23.83 & 31.84 \\

    & {\method}-s & \textbf{84.69} & 12.13 & 21.23 & \textbf{70.69} & 11.99 & 20.50 \\
    
    & {\method} & \underline{71.24} & \textbf{61.55} & \textbf{66.04} & \underline{51.60} & \textbf{47.22} & \textbf{49.31} \\

    \bottomrule
    \multicolumn{8}{l}{P: Precision, R: Recall.}
    
  \end{tabular}
  \label{tab:abstudy}
  \end{center}
\end{table}

\subsection{RQ4: Efficiency of {\method}}\label{subsec:efficiency}
We compare the detection time of {\method} with other TPL detection tools on datasets $D_1$ and $D_2$. As shown in Table~\ref{tab:time}, we record the total processing time for each app by the TPL detection tools, and we compute the first quartile (Q1), median, third quartile (Q3), and mean of the detection times. We find that {\method} is slower than LibHunter, LIBLOOM and LibScan. The reason for LIBLOOM's superior speed is its use of a two-stage bloom filter, which provides highly scalable TPL detection by converting extracted features to hash values, thereby significantly accelerating the matching phase. LibScan, on the other hand, is somewhat slower than LIBLOOM due to the time spent in opcode extraction and matching.

{\method} is relatively slower due to two primary reasons. First, {\method} spends considerable time in the candidate app class list generation step, with the feature propagation and matching of node features in graph structures being the main time-consuming phases. Second, {\method}'s precise version-level detection introduces additional time overhead by generating the class summaries of call sequences. Despite LibHunter, LibScan and LIBLOOM having less detection times compared to {\method}, they exhibit inferior performance in version-level$^{\dagger}$ detection performance.
Considering the significance of TPL detection, we deem it reasonable for {\method} to achieve better performance at the expense of certain efficiency.

\begin{table}[htbp]
  \caption{Average detection efficiency (s) of different tools on $D_1$ and $D_2$ (453 TPLs in total).}
    \small
  \begin{center}
  \begin{tabular}{cccccc}
    \toprule
    & LibPecker & LibHunter & LIBLOOM & LibScan & {\method}  \\
    \hline
    
    \textbf{Q1} & 2,270.04 & 35.00 & 2.00 & 45.00 & 15.39 \\

    \textbf{mean} & 2,442.74 & 63.95 & 5.00 & 46.05 & 129.78 \\

    \textbf{median} & 2,546.37 & 49.00 & 2.56 & 49.00 & 109.68 \\

    \textbf{Q3} & 2,779.70 & 75.00 & 3.98 & 55.00 & 192.24 \\

    \bottomrule
    
  \end{tabular}
  \label{tab:time}
  \end{center}
\end{table}

\section{Discussion}

\noindent \textbf{Limitations and Future Work.}
{\method} cannot handle all types of obfuscations. For advanced obfuscation techniques such as reflective invocation, \texttt{Dex} file encryption, code virtualization, and others ~\cite{obfuscapk}, the reliability of {\method}'s class matching step will be compromised, resulting in TPL detection failures. Additionally, like previous approaches, {\method} tunes its thresholds separately on different datasets for evaluation. However, in real-world scenarios where some apps with mixed obfuscation and optimization statuses coexist, it remains challenging for {\method} to identify an optimal threshold setting to achieve high performance.

LibHunter~\cite{libhunter} is specifically designed for optimized apps and achieves better TPL detection performance on optimized datasets compared to {\method}, but performs worse on obfuscated datasets. In real-world scenarios, it is often unclear whether the input app has undergone obfuscation or optimization. Therefore, it is necessary for TPL detection tools to cover obfuscated and optimized apps. However, as noted in the types of optimization listed by LibHunter~\cite{libhunter}, the code changes introduced by optimization may conflict with the principles of TPL detection tools to handle obfuscation. For example, to address dead code removal, {\method} tolerates cases where the code in an app’s class is less than that in the corresponding TPL class, whereas method inlining optimization inflates the code within app classes. Therefore, finding a balance between handling obfuscation and optimization is inherently challenging, and we leave it to future work to address.

Although we believe that {\method} achieves a balance between effectiveness and efficiency, there remains room for improvement in detection efficiency compared to state-of-the-art baseline tools. The outstanding scalability of LIBLOOM stems from its reliance solely on class-level structural features and its efficient detection mechanism based on bloom filter design, making it suitable for high-throughput scenarios. However, as shown in \S\ref{subsec:classlevel}, the hashing of features inevitably leads to collisions, compromising the class-level detection performance of LIBLOOM and reducing reliability. In contrast, both LibScan and LibHunter employ multi-process implementations to fully utilize CPU resources for accelerating the detection process, offering an effective approach to enhance the efficiency of {\method}.


\noindent \textbf{Threats to validity.}
{\method} leverages the structural information of CDGs to generate candidate class lists and assesses structural similarity of matched nodes during TPL detection to identify specific versions. This design exhibits resilience to code obfuscation, as class dependency relationships within a complete functional module are typically unaffected by obfuscation. The class matching module, based on class summary similarity, extracts parameter-relevant instruction slices for member matching and generates class summaries using field operations, mitigating the impact of obfuscations like control flow flattening. 
However, in optimized apps employing method inlining, the inlined methods (callees) may fail to match, and the inflated code at the unique invocation site (caller) could lead to member matching failures, thereby affecting class matching results. Furthermore, {\method} demonstrates limited resilience to more advanced obfuscation techniques, such as method overloading and reflection invocations~\cite{obfuscapk}. Nonetheless, the applicability of these advanced obfuscation techniques is constrained, as they impose significant overhead to maintain app functionality and performance.

Furthermore, {\method} is intentionally crafted to identify Java libraries (JAR) as well as Android libraries (AAR). Consequently, its generalizability to native libraries within apps may be limited, as native libraries may employ complex obfuscation techniques that {\method} does not account for (e.g., code virtualization~\cite{8732327}). However, code virtualization inevitably incurs the overhead of virtual machine interpretation and execution, making it typically used to protect critical functions. Existing research on deobfuscating virtualized code~\cite{10.1007/978-3-319-89500-0_28,10.1145/2046707.2046739} may contribute to the detection of native libraries.





\section{Conclusion}
We proposed {\method}, a class structural similarity-based version-level TPL detection tool. {\method} generates candidate app class lists for TPL classes using the feature similarity of nodes in CDGs associated with the app and TPL and then performs class matching based on the similarity of class functionality summary. Finally, {\method} achieves version-level TPL detection by identifying structural similarity between the sub-graphs formed by matched classes within the app and TPL CDG. Experimental results show that {\method} outperforms baseline tools in both library-level and version-level$^{\dagger}$ detection, achieving F1 scores of 97.64\% and 84.82\% on obfuscated dataset, respectively, demonstrating its effectiveness. Additionally, the superior performance of {\method} in class-level detection underscores its reliability, making it well-suited for downstream tasks that depend on specific TPL code. 
The source code of the {\method} and the experimental results are available at \url{https://zenodo.org/records/15238860}.
\begin{acks}
This paper is supported by the National Natural Science Foundation of China under No.62202457 and the Open Source Community Software Bill of Materials (SBOM) Platform under No.E3GX310201. This paper is also supported by YuanTu Large Research Infrastructure.
\end{acks}

\bibliographystyle{ACM-Reference-Format}
\bibliography{ref}




\end{document}